\documentstyle[aps,preprint]{revtex}

\begin{document}
\title{Scaling theory of magnetic ordering in the Kondo lattices with anisotropic
exchange interactions}
\author{V.Yu. Irkhin$^{*}$ and M.I. Katsnelson}
\address{Institute of Metal Physics, 620219 Ekaterinburg, Russia}
\maketitle

\begin{abstract}
The lowest-order scaling consideration of the magnetic state formation in
the Kondo lattices is performed within the $s-f$ model with inclusion of
anisotropy for both the $f-f$ coupling and $s-f$ exchange interaction. The
Kondo renormalizations of the effective transverse and longitudinal $s-f$
coupling parameters, spin-wave frequency, gap in the magnon spectrum and
ordered moment are calculated in the case of both ferro- and
antiferromagnets. The anisotropy-driven change of the scaling behavior
(e.g., critical value of $g$ for entering the strong-coupling region and the
corresponding critical exponents) is investigated numerically for $N=2$ and
analytically in the large-$N$ limit. The dependence of the effective Kondo
temperature on the bare $s-f$ coupling parameter $g$ weakens in the presence
of anisotropy. The relative anisotropy parameters for both the $s-f$ and $%
f-f $ coupling are demonstrated to decrease during the renormalization
process. The role of next-nearest exchange interactions for this effect in
the antiferromagnet is discussed.
\end{abstract}


\section{Introduction}

Anomalous rare earth and actinide compounds, including so-called Kondo
lattices and heavy-fermion systems, are studied extensively by both
experimentalists and theorists\cite{Stewart,Adr,Lac,Fulde,Zw}. It is now a
common point of view that the most interesting peculiarities of electronic
and magnetic propeties of these systems are due to the interplay of the
on-site Kondo effect and intersite magnetic interactions. Whereas the
one-impurity Kondo problem, being itself very difficult and rich, is now
studied in detail\cite{Tsv}, the Kondo-lattice problem is still a subject of
many investigations\cite{Lac,Fulde,Zw,kondo}. Usually this problem is
studied within the standard $s-f$ exchange or Anderson models. On the other
hand, strong effects of crystal field and anisotropic interactions are
expected in anomalous $4f$ and $5f$-systems (see, e.g.,\cite{FLow}). These
effects can lead to anisotropic terms in the Hamiltonian. It is well known
that the change of symmetry of the $s-f$ exchange interaction modifies
qualitatively the infrared behavior in the one-impurity case\cite
{Noz,Tsv,ikt}. Thus one could expect that similar effects should take place
in the lattice case. Therefore a question arises whether anisotropic
contributions are important also in the problem of competition of the Kondo
effect and magnetism. It should be noted that this question is relevant not
only for magnetic systems, but also for models with pseudospin degrees of
freedom (e.g., for strongly anharmonic crystals demonstrating band
Jahn-Teller effect\cite{ikt1}).

Theoretical investigations of the Kondo-lattice problem use as a rule
methods appropriate for calculating low-temperature properties in the
strong-couping regime ($1/N$-expansion\cite{Bick}, slave-boson technique).
However, these methods are not convenient for the description of the
transition to the weak-coupling regime (in particular, even derivation of
the standard Kondo logaruithms is here a non-trivial problem\cite{Col}). In
our previous paper\cite{kondo} we have proposed an alternative approach
which starts from the weak-coupling regime and is based on summing up
leading divergent terms by the renormalization group method. We have
investigated the formation of magnetic state in the periodic $s-f$ exchange
and Coqblin-Schrieffer models with the $f$-subsystem being described by the
isotropic Heisenberg Hamiltonian. The aim of the present paper is the
investigation of formation of the magnetic Kondo-lattice state for various
magnetic phases with account of the anisotropy in both the localized-spin
subsystem and $s-f$ exchange interaction.

In Sect.2 we discuss the theoretical model and calculate the logarithmic
Kondo corrections to the spin-wave spectrum of anisotropic metallic ferro-
and antiferromagnets. In Sect.3 we derive the lowest-order scaling equations
for the effective transverse and longitudinal $s-f$ exchange parameters and
renormalized magnon frequencies. In Sect.4 we present a simple analytical
solution with magnon spectrum renormalizations being neglected, which is
possible in the large-$N$ limit of the Coqblin-Schrieffer model. In Sect.5
we discuss results of the numerical solution of the full scaling equations
for $N=2$  in the presence of the anisotropy in localized-spin system only
and investigate new features which occur in comparison with the isotropic
case. In Sect.6, influence of the anisotropic $s-f$ coupling is considered.

\section{Theoretical models and Kondo corrections to the spectrum of spin
excitations}

To treat the Kondo effect in a lattice we use the $s-d(f)$ exchange model 
\begin{equation}
H=\sum_{{\bf k}\sigma }t_{{\bf k}}c_{{\bf k}\sigma }^{\dagger }c_{{\bf k}%
\sigma }^{}+H_f+H_{sf}=H_0+H_{sf}  \label{H}
\end{equation}
where $t_{{\bf k}}$ is the band energy. We consider the pure spin $s-d(f)\,$%
exchange Hamiltonian with 
\begin{eqnarray}
H_f &=&\sum_{{\bf q}}J_{{\bf q}}{\bf S}_{{\bf -q}}{\bf S}_{{\bf q}}+\eta
\sum_{{\bf q}}J_{{\bf q}}S_{-{\bf q}}^zS_{{\bf q}}^z-K\sum_i(S_i^z)^2,
\label{sfe} \\
\,\,H_{sf} &=&-\sum_{{\bf kk}^{\prime }\alpha \beta }[I_{\parallel }S_{{\bf %
k-k}^{\prime }}^z(c_{{\bf k}\uparrow }^{\dagger }c_{{\bf k}^{\prime
}\uparrow }^{}-c_{{\bf k}\downarrow }^{\dagger }c_{{\bf k}^{\prime
}\downarrow }^{})+I_{\perp }(S_{{\bf k-k}^{\prime }}^{+}c_{{\bf k\downarrow }%
}^{\dagger }c_{{\bf k}^{\prime }\uparrow }^{}+S_{{\bf k-k}^{\prime }}^{-}c_{%
{\bf k}\uparrow }^{\dagger }c_{{\bf k}^{\prime }\uparrow }^{})]  \label{sfh}
\end{eqnarray}
where ${\bf S}_i$ and ${\bf S}_{{\bf q}}$ are spin operators and their
Fourier transforms, $\eta >0$ and $K>0$ are the parameters of the two-site
and single-site easy-axis magnetic anisotropy in the $f$-subsystem,
respectively. Note that our consideration can be formally generalized on the
Coqblin-Schrieffer model with arbitrary $N$ (cf. Ref.\cite{kondo}) or to a
more general form of the $s-f$ coupling parameter matrix\cite{ikt}. For
simplicity, we neglect {\bf k}-dependence of the $s-f$ parameter which
occurs in the degenerate $s-f$ model due to the angle dependence of the
coupling (see Ref.\cite{kondo}).Of course, in fact the $f-f$ exchange has
usually the Ruderman-Kittel-Kasuya-Yosida (RKKY) origin and is determined by
the same $s-f$ coupling. Thus, generally speaking, the anisotropy of the $s-f
$ coupling and $f$-subsystem are not independent. However, crystal-field
effects are known to be more important in formation of magnetic anisotropy
in rare-earth metals than anisotropic exchange interactions\cite{irk} (in
this case, the anisotropic $s-f$ coupling is obtained by expansion in the
parameter $k_Fr_f,$ $r_f$ being the $f$-shell radius, and contains, unlike (%
\ref{sfh}), terms of the type $({\bf kS})({\bf k}^{\prime }{\bf S})$). On
the other hand, the situation, where anisotropy occurs in the $s-f$ coupling
only, may be also considered: this corresponds to the strong ``direct'' $f-f$
exchange (e.g., superexchange) interaction which is characteristic for some $%
f$-compounds. In the Coqblin-Schrieffer model, which is more appropriate for
cerium compounds, crystal field results in occurrence of anisotropic $s-f$
coupling\cite{Maekawa,Cox} and new excitation branches\cite{FLow,kondo}. For
simplicity, we restrict ourselves to treatment of a single magnon mode in
the simplest $s-f$ model (\ref{H}).

In the ferromagnetic (FM) state the spin-wave spectrum for the Hamiltonian (%
\ref{sfe}) reads 
\begin{eqnarray}
\omega _{{\bf q}} &=&\omega _0+\omega _{ex}({\bf q}),  \label{wf} \\
\omega _{ex}({\bf q}) &=&2S(J_{{\bf q}}-J_0),\omega _0=-2S\eta J_0+(2S-1)K
\end{eqnarray}
To find the Kondo logarithmic corrections to the spin-wave spectrum we
calculate the magnon Green's functions in the model (\ref{H}). For a
ferromagnet we obtain to second order in $I$ (cf. the calculations in the
isotropic case\cite{Aus}) 
\begin{eqnarray}
\langle \langle b_{{\bf q}}|b_{{\bf q}}^{\dagger }\rangle \rangle _\omega
&=&\left[ \omega -\omega _{{\bf q}}\right. -2\sum_{{\bf p}}(J_0+J_{{\bf q-p}%
}-J_{{\bf p}}-J_{{\bf q}}+\omega _0/2S)\langle b_{{\bf p}}^{\dagger }b_{{\bf %
p}}\rangle  \nonumber \\
&&\ \ \ -2S\sum_{{\bf k}}\left( I_{\perp }^2\frac{n_{{\bf k}}-n_{{\bf k-q}}}{%
\omega +t_{{\bf k}}-t_{{\bf k-q}}}-I_{\parallel }^2\frac{\partial n_{{\bf k}}%
}{\partial t_{{\bf k}}}\right) \left. -2\sum_{{\bf p}}(I_{\parallel }^2\phi
_{{\bf pq}\omega }-I_{\perp }^2\phi _{{\bf p}00})\right] ^{-1}  \label{FGf}
\end{eqnarray}
where we have taken into account kinematic requirements in the magnon
anharmonicity terms by introducing the factor of $(2S-1)/2S$ at $K$ (this
replacement may be justified by considering higher-order terms in the formal
parameter $1/S$), $n_{{\bf k}}=n(t_{{\bf k}})$ is the Fermi function, 
\begin{equation}
\phi _{{\bf pq}\omega }=\sum_{{\bf k}}\frac{n_{{\bf k}}(1-n_{{\bf k+p-q}})}{%
\omega +t_{{\bf k}}-t_{{\bf k+p-q}}-\omega _{{\bf p}}}  \label{Fif}
\end{equation}
The averages that enter (\ref{FGf}) can be obtained from the spectral
representation for the Green's function (\ref{FGf}) to first order in $1/S$
and contain the singular contributions 
\begin{equation}
\delta \langle b_{{\bf q}}^{\dagger }b_{{\bf q}}\rangle =SI_{\perp }^2\Phi _{%
{\bf q}}  \label{bbf}
\end{equation}
with 
\begin{equation}
\Phi _{{\bf q}}^{FM}=\sum_{{\bf k}}\frac{n_{{\bf k}}(1-n_{{\bf k+q}})}{(t_{%
{\bf k}}-t_{{\bf k+q}}-\omega _{{\bf q}})^2}  \label{Fifm}
\end{equation}
Expanding the denominators of (\ref{Fif}) in the magnon frequencies we
obtain the singular correction to the pole of the magnon Green's function 
\[
\delta \omega _{{\bf q}}=-2S(I_{\perp }^2+I_{\parallel }^2)\sum_{{\bf p}}(J_{%
{\bf p}}-J_{{\bf q-p}}+J_{{\bf q}}-J_0+\omega _0/2S)\Phi _{{\bf p}}^{FM} 
\]
This result can be represented as 
\begin{eqnarray}
\delta \omega _{ex}({\bf q})/\omega _{ex}({\bf q}) &=&-(I_{\perp
}^2+I_{\parallel }^2)(1-\alpha _{{\bf q}})\sum_{{\bf p}}\Phi _{{\bf p}}^{FM}
\label{dwf} \\
\delta \omega _0/\omega _0 &=&-(I_{\perp }^2+I_{\parallel }^2)\sum_{{\bf p}%
}\Phi _{{\bf p}}^{FM}  \label{dw0}
\end{eqnarray}
where $0<\alpha _{{\bf q}}<1.$ Passing into real space (see \cite{kondo})
yields 
\begin{equation}
\alpha _{{\bf q}}=\sum_{{\bf R}}J_{{\bf R}}\left| \langle e^{i{\bf kR}%
}\rangle _{t_{{\bf k}}=E_F}\right| ^2[1-\cos {\bf qR]/}\sum_{{\bf R}}J_{{\bf %
R}}[1-\cos {\bf qR],}  \label{Alpq}
\end{equation}
In the approximation of nearest neighbors at the distance $d,$ the quantity $%
\alpha $ does not depend on ${\bf q}$. For a spherical Fermi surface we have 
\begin{equation}
\alpha _{{\bf q}}=\alpha =\left| \langle e^{i{\bf kR}}\rangle _{t_{{\bf k}%
}=E_F}\right| ^2=\left( \frac{\sin k_Fd}{k_Fd}\right) ^2  \label{ALp}
\end{equation}
Hereafter we put $\alpha =$ const. Then we may use in further consideration
of the scaling equations a single renormalization parameter, rather than the
whole function of {\bf q}.

Now we consider a two-sublattice antiferromagnet (AFM) with the wavevector
of magnetic structure {\bf Q, } 
\[
\langle S_i^z\rangle =S\cos {\bf QR}_i,\,\langle \,S_i^y\rangle =\,\langle
S_i^x\rangle =0
\]
($J_{{\bf Q}}=J_{\min }<0;$ 2{\bf Q }is equal to a reciprocal lattice
vector, so that $\cos ^2{\bf QR}_i=1,\,\sin ^2{\bf QR}_i=0;$ only in this
case we can retain the definitions of $I_{\perp }$ and $I_{\parallel }$ in
the local coordinate system). Passing to the spin-deviation operators in the
local coordinate system where 
\begin{equation}
S_i^z=\hat S_i^z\cos {\bf QR}_i,\,S_i^y=\hat S_i^y\cos {\bf QR}_i,\,S_i^x=%
\hat S_i^x
\end{equation}
we derive 
\begin{eqnarray}
H_f &=&\text{const }+\sum_{{\bf q}}[C_{{\bf q}}b_{{\bf q}}^{\dagger }b_{{\bf %
q}}+\frac 12D_{{\bf q}}(b_{-{\bf q}}b_{{\bf q}}+b_{{\bf q}}^{\dagger }b_{-%
{\bf q}}^{\dagger })]+...  \label{Hsw} \\
C_{{\bf q}} &=&S(J_{{\bf Q+q}}+J_{{\bf q}}-2J_{{\bf Q}}(1+\eta
))+(2S-1)K,\,D_{{\bf q}}=S(J_{{\bf q}}-J_{{\bf Q+q}})
\end{eqnarray}
Diagonalizing (\ref{Hsw}) we obtain the spin-wave spectrum 
\begin{eqnarray}
\omega _{{\bf q}}^2 &=&C_{{\bf q}}^2-D_{{\bf q}}^2\simeq \omega _0^2+\omega
_{ex}^2({\bf q})  \label{waf} \\
\omega _{ex}({\bf q}) &=&2S(J_{{\bf q}}-J_{{\bf Q}})^{1/2}(J_{{\bf Q+q}}-J_{%
{\bf Q}})^{1/2}, \\
\omega _0^2 &=&2S(J_0-J_{{\bf Q}})[(2S-1)K-2S\eta J_{{\bf Q}}]
\end{eqnarray}
where we have neglected a weak wavevector dependence of $\omega _0.$

The Kondo correction to the spectrum (\ref{waf}) reads (cf. Refs.\cite
{kondo,IK96}) 
\begin{eqnarray}
\delta \omega _{{\bf q}}^2\ &=&-2\sum_{{\bf p}}[I_{\perp }^2\omega _{{\bf q}%
}^2+2S^2I_{\perp }^2(J_{{\bf Q+q}}+J_{{\bf q}}-2J_{{\bf Q}})(J_{{\bf p}}+J_{%
{\bf Q+p}}-J_{{\bf Q+q-p}}-J_{{\bf q-p}})  \label{dw2} \\
&&\ \ +2(I_{\parallel }^2-I_{\perp }^2)(C_{{\bf q}}C_{{\bf p-q}}-D_{{\bf q}%
}D_{{\bf p-q}})]\Phi _{{\bf p}}^{AFM}  \nonumber
\end{eqnarray}
with 
\begin{equation}
\Phi _{{\bf q}}^{AFM}=\sum_{{\bf k}}\frac{n_{{\bf k}}(1-n_{{\bf k+q}})}{(t_{%
{\bf k}}-t_{{\bf k+q}})^2-\omega _{{\bf q}}^2}  \label{Fiafm}
\end{equation}
For an antiferromagnet in the nearest-neighbor approximation ($J_{{\bf Q+q}%
}=-J_{{\bf q}}$) we obtain from (\ref{dw2}) 
\begin{eqnarray}
\delta \omega _{ex}^2({\bf q})/\ \omega _{ex}^2({\bf q})\ &=&-2[I_{\perp
}^2-\alpha (I_{\parallel }^2-I_{\perp }^2)]\sum_{{\bf p}}\Phi _{{\bf p}%
}^{AFM}  \label{dw0e} \\
\ \delta \omega _0^2/\omega _0^2 &=&-2[I_{\perp }^2\ -(1-\alpha
)s(I_{\parallel }^2-I_{\perp }^2)]\sum_{{\bf p}}\Phi _{{\bf p}}^{AFM} 
\nonumber
\end{eqnarray}
where $s=4S^2J_{{\bf Q}}^2/\omega _0^2\gg 1.$

Introducing next-nearest-neighbor interactions and putting for simplicity $%
I_{\parallel }^{}=I_{\perp }^{}=I$ we obtain 
\begin{equation}
\ \delta \omega _{{\bf q}}^2\ =-2[\omega _{{\bf q}}^2-8S^2\alpha _{{\bf q}%
}^{(2)}(J_{{\bf q}}^{(2)}-J_{{\bf Q}})(J_{{\bf q}}^{(2)}-J_{{\bf Q}%
}^{(2)})]\sum_{{\bf p}}\Phi _{{\bf p}}^{AFM}  \label{dw22}
\end{equation}
where 
\[
J_{{\bf q}}^{(1,2)}=\frac 12(J_{{\bf q}}\mp J_{{\bf q+Q}}) 
\]
correspond to the contribution of nearest and next-nearest neighbors, and $%
\alpha _{{\bf q}}^{(2)}$ is given (\ref{Alpq}) with the sum over the
next-nearest neighbors. Provided that next-nearest neighbor exchange
interaction is ferromagnetic (so that the AFM ordering is stable), $J_{{\bf q%
}}^{(2)}-J_{{\bf Q}}^{(2)}>0$ and the next-nearest neighbors result in
decreasing the Kondo suppression of the magnon frequency, as well as nearest
neighbors in the FM case. Then, instead of (\ref{dw0e}), we can use
phenomenologically (e.g., in the long-wave limit) the expression 
\begin{eqnarray}
\ \delta \omega _{ex}^{}({\bf q})/\ \omega _{ex}^{}({\bf q})\ &=&-(1-\alpha
^{\prime })I^2\sum_{{\bf p}}\Phi _{{\bf p}}^{AFM}  \label{dw'} \\
\delta \omega _0^{}/\ \omega _0 &=&-I^2\sum_{{\bf p}}\Phi _{{\bf p}}^{AFM} 
\nonumber
\end{eqnarray}
with $\alpha ^{\prime }\propto \alpha ^{(2)}J^{(2)}/J^{(1)}.$ In the
opposite case of AFM next-nearest neighbor exchange the situation is more
complicated. In particular, the simple collinear antiferromagnetism can
become unstable, and formation of the spiral structure is possible. Note the
difference between the FM and AFM cases by a factor of 2, which is due to
violation of time-reversal symmetry in a ferromagnet (terms that are linear
in $\omega $ give a contribution).

The quantities (\ref{Fifm}), (\ref{Fiafm}) determine also the singular
correction to the (sublattice) magnetization 
\begin{equation}
\delta \bar S/S=-\frac 1S\sum_{{\bf q}}\delta \langle b_{{\bf q}}^{\dagger
}b_{{\bf q}}\rangle =-I_{\perp }^2\sum_{{\bf q}}\Phi _{{\bf q}}^{FM,AFM}
\label{Ssf}
\end{equation}

\section{Scaling equations}

Using the perturbation theory results we can write down the system of
scaling equations in the case of the Kondo lattice for various magnetic
phases. Their derivation in the isotropic case is described in detail in Ref.%
\cite{kondo}. As well as in this paper, we apply the ``poor man scaling''
approach \cite{And}. In this method one considers the dependence of
effective (renormalized) model parameters on the cutoff parameter $C$ ($%
-D<C<0,$ here and hereafter the energy is calculated from the Fermi energy $%
E_F=0$) which occurs at picking out the Kondo singular terms.

The renormalization of $I_{\parallel }$ is obtained from renormalization of
the magnetic splitting in electron spectrum, and of $I_{\perp }^{}$ from
renormalization of the second-order contribution to the electron self-energy
(see corresponding perturbation expressions for a ferromagnet in Ref.\cite
{Aus1}). The renormalized $I_{\perp }^{}$ chosen in such a way coincides
with the three-leg vertex (with two electron lines with opposite spins and
one magnon line) which yields the most natural definition in a magnetically
ordered case and agrees with the one-impurity scaling consideration\cite{Tsv}
To find the equation for the effective coupling parameters $I_{ef}^\alpha
(C) $ ($I_{ef}^\alpha (-D)=I_{ef}^\alpha $) we have to calculate the
contribution of intermediate electron states near the Fermi level with $C<t_{%
{\bf k+q}}<C+\delta C$ in the sums that enter expressions for the
self-energies (which include, unlike Ref.\cite{kondo}, magnon frequencies
with a gap). Then we obtain 
\begin{eqnarray}
\delta I_{ef}^{\parallel }(C) &=&2\rho I_{\perp }^2\eta (-\frac{\overline{%
\omega }_{ex}}C,-\frac{\omega _0}C)\delta C/C  \label{ief} \\
\delta I_{ef}^{\perp }(C) &=&2\rho I_{\perp }^{}I_{\parallel }\eta (-\frac{%
\overline{\omega }_{ex}}C,-\frac{\omega _0}C)\delta C/C  \nonumber
\end{eqnarray}
where $\rho $ is the density of states at the Fermi level, $\overline{\omega 
}_{ex}$ is a characteristic spin-fluctuation energy, $\omega _0$ is the gap
in the spin-wave spectrum, $\eta (x)$ is a scaling function which satisfies
the condition $\eta (0)=1$ which guarantees the correct one-impurity limit.
For both FM and AFM phases we have 
\begin{equation}
\eta ^{FM,AFM}(-\frac{\overline{\omega }_{ex}}C,-\frac{\omega _0}C)=\langle
(1-\omega _{{\bf k-k}^{\prime }}^2/C^2)^{-1}\rangle _{t_k=t_{k^{\prime
}}=E_F}  \label{etafm}
\end{equation}
where the magnon frequencies are given by (\ref{wf}), (\ref{waf}). The
corresponding analytical expressions are presented in Appendix.

The $C$-dependent renormalizations of the spin-wave frequencies and
ground-state moment are obtained in the same way as in the isotropic case%
\cite{kondo} from (\ref{dwf}), (\ref{dw2}) (\ref{dw0e}), (\ref{dw'}), (\ref
{Ssf}) and expressed in terms of the same scaling functions. Introducing the
dimensionless coupling constants 
\[
g_{ef}^\alpha (C)=-2\rho I_{ef}^\alpha (C),\,g_\alpha =-2\rho I_\alpha 
\]
(we will drop sometimes the index $\parallel ,$ but not $\perp ,$ so that $%
g_{ef}^{}(C)\equiv g_{ef}^{\parallel }(C)$) and replacing $g_\alpha
\rightarrow g_{ef}^\alpha (C),\,\overline{\omega }_{ex}\rightarrow \overline{%
\omega }_{ex}(C),$ $\omega _0\rightarrow \omega _0(C)$ in the right-hand
parts of (\ref{ief}) and expressions for $\delta \overline{\omega }%
_{ex}(C),\delta \omega _0(C)$ and $\delta \overline{S}_{ef}(C),$ we obtain
the system of scaling equation 
\begin{eqnarray}
\partial g_{ef}^{\parallel }(C)/\partial C &=&-[g_{ef}^{\perp }(C)]^2\Lambda
\label{gl} \\
\partial g_{ef}^{\perp }(C)/\partial C &=&-g_{ef}^{\parallel
}(C)g_{ef}^{\perp }(C)\Lambda  \label{glpp} \\
\partial \ln \overline{\omega }_{ex}(C)/\partial C &=&a\Lambda /2\times
\left\{ 
\begin{tabular}{ll}
$\{[g_{ef}^{\parallel }(C)]^2+[g_{ef}^{\perp }(C)]^2\}/2$ & FM \\ 
$\lbrack g_{ef}^{\perp }(C)]^2-\alpha \{[g_{ef}^{\parallel
}(C)]^2-[g_{ef}^{\perp }(C)]^2\}$ & AFM
\end{tabular}
\right.  \label{wl} \\
\partial \ln \omega _0(C)/\partial C &=&b\Lambda /2\times \left\{ 
\begin{tabular}{ll}
$\{[g_{ef}^{\parallel }(C)]^2+[g_{ef}^{\perp }(C)]^2\}/2$ & FM \\ 
$\lbrack g_{ef}^{\perp }(C)]^2-s(1-\alpha )\{[g_{ef}^{\parallel
}(C)]^2-[g_{ef}^{\perp }(C)]\}$ & AFM
\end{tabular}
\right.  \label{w0} \\
\partial \ln \overline{S}_{ef}(C)/\partial C &=&[g_{ef}^{\perp
}(C)]^2\Lambda /2  \label{sl}
\end{eqnarray}
where 
\begin{equation}
\Lambda =\Lambda (C,\overline{\omega }_{ex}(C),\omega _0(C))=\eta (-%
\overline{\omega }_{ef}(C)/C,-\omega _0(C)/C)/C,
\end{equation}
and 
\begin{equation}
a=\left\{ 
\begin{array}{cc}
2(1-\alpha ) & \text{FM} \\ 
1-\alpha ^{\prime } & \text{AFM}
\end{array}
\right. ,\,b=\left\{ 
\begin{array}{cc}
2 & \text{FM} \\ 
1 & \text{AFM}
\end{array}
\right.
\end{equation}

The integral of motion of the system (\ref{gl}), (\ref{glpp}) reads 
\begin{equation}
\lbrack g_{ef}^{\parallel }(C)]^2-[g_{ef}^{\perp }(C)]^2=\mu ^2=g_{\parallel
}^2-g_{\perp }^2=\text{const}
\end{equation}
so that the equation (\ref{gl}) takes the form 
\begin{equation}
\partial g_{ef}^{}(C)/\partial C=-[g_{ef}^2(C)-\mu ^2]\Lambda  \label{gmu}
\end{equation}

\section{Analytical solution in the large-$N$ limit}

In the formal large-$N$ limit in the Coqblin-Schrieffer model where $%
2\rightarrow N$ in Eqs.(\ref{wl})-(\ref{sl})) we can neglect
renormalizations of magnon frequencies (note that the true large-$N$ limit
in the FM case is somewhat different since symmetry of spin-up and spin-down
state contributions is violated for $N>2$, see Ref.\cite{kondo}). Note that
the same approximation is valid for $N=2$ provided that $g$ is well below
the critical value $g_c$ for entering the strong-coupling region. Then, on
taking into account (\ref{eta1}),(\ref{eta3}), equation (\ref{gmu}) can be
integrated analytically to obtain 
\begin{equation}
\frac 1\mu [%
\mathop{\rm arctanh}
(\mu /g_{ef}(C))-%
\mathop{\rm arctanh}
(\mu /g)]=G(C)=-\int_{-D}^C\frac{dC^{\prime }}{C^{\prime }}\eta (-\frac{%
\overline{\omega }}{C^{\prime }},-\frac{\omega _0}{C^{\prime }})
\label{g111}
\end{equation}
\begin{eqnarray}
G^{FM}(C) &=&\ln |C/D|-((1+w)/2)[(C/\overline{\omega }-1)\ln |1-\overline{%
\omega }/C|-(C/\overline{\omega }+1)\ln |1+\overline{\omega }/C|]  \label{g2}
\\
&&\ \ \ \ \ \ \ \ \ \ \ \ \ \ \ \ \ +(w/2)[(C/\omega _0-1)\ln |1-\omega
_0/C|-(C/\omega _0+1)\ln |1+\omega _0/C|]-1  \nonumber \\
G_{d=3}^{AFM}(C) &=&\ln |C/D|-\frac 12[(1+w^2)(C^2/\overline{\omega }%
^2-1)\ln |1-\overline{\omega }^2/C^2|  \label{g3} \\
&&\ \ \ \ \ \ \ \ \ \ \ \ \ \ \ \ \ -w^2(C^2/\omega _0^2-1)\ln |1-\omega
_0^2/C^2|+1]  \nonumber \\
G_{d=2}^{AFM}(C) &=&[\theta (C^2-\overline{\omega }^2)+\theta (\omega
_0^2-C^2)]\ln (\frac 12(\sqrt{|C^2-\omega _0^2|}+\sqrt{|C^2-\overline{\omega 
}^2|})/D)  \label{g4} \\
&&\ \ \ \ \ \ \ \ \ \ \ \ \ \ \ \ \ +\theta (C^2-\omega _0^2)\theta (%
\overline{\omega }^2-C^2)\ln (\overline{\omega }_{ex}/2D)  \nonumber
\end{eqnarray}
where 
\begin{equation}
\,\overline{\omega }=\left\{ 
\begin{array}{cc}
\omega _0+\overline{\omega }_{ex} & \text{FM} \\ 
\sqrt{\omega _0^2+\overline{\omega }_{ex}^2} & \text{AFM}
\end{array}
\right.   \label{wtot}
\end{equation}
The scaling trajectories described by (\ref{g111})-(\ref{g4}) are shown in
Fig.1 for $\mu \neq $ $0,w=0$ and in Fig.2 for $w\neq 0,\mu =0$. Note that
these pictures describe adequately the case $N=2,$ since $g=0.15$ is
considerably lower than the critical values.

The anisotropy of $s-f$ coupling results in that the dependence $%
1/g_{ef}(\xi )$ becomes non-linear at small $\xi =\ln |D/C|$ where the
one-impurity behavior takes place 
\begin{equation}
1/g_{ef}(\xi )\simeq \mu \tanh [%
\mathop{\rm arctanh}
(\mu /g)-\mu \xi ]
\end{equation}
However, this non-linearity is not too strong even for $\mu /g=2/3$ (Fig.1).
Of course, the curves with $\mu \neq 0$ go considerably higher, since the
bare coupling parameter $g_{\perp }$ decreases with $\mu .$

For $w\neq $ $0$ the function $1/g_{ef}(\xi )$ has a minimum both in the AFM
and FM cases, position of which, $C_{\min }$, is determined by (\ref{min1})-(%
\ref{min3}) (in the isotropic case, the minimum occurs in the 3D AFM case
only). The minimum may result in non-monotonic temperature dependences of
physical quantities which are sensitive to the Kondo effect, e.g., of the
effective magnetic moment. The depth of the minimum $\Delta =G_{\min }-G(0)$
is given by 
\begin{equation}
\Delta =\left\{ 
\begin{array}{cc}
\frac 12[(1+w)\ln (1+w)-w\ln w] & \text{FM} \\ 
\frac 12[\ln 2+(1+w^2)\ln (1+w^2)-w^2\ln w^2] & \text{3D AFM} \\ 
\ln (w+\sqrt{1+w^2}) & \text{2D AFM}
\end{array}
\right.  \label{delta}
\end{equation}
Note that at $w\gg 1$ we have in all the cases $\Delta \propto \ln w.$

The critical value $g_c$ is determined by the condition 
\begin{equation}
1/g_c=-(1/\mu )\tanh (\mu G_{\min })\simeq |G_{\min }-(\mu ^2/3)G_{\min
}^3|,\mu \ll 1  \label{gcmu}
\end{equation}
where 
\begin{equation}
-G_{\min }=\left\{ 
\begin{array}{cc}
\lambda +1+\frac 12[w\ln w+(1-w)\ln (1+w)] & \text{FM} \\ 
\lambda +\frac 12[1+\ln 2+\ln (1+w^2)] & \text{3D AFM} \\ 
\lambda & \text{2D AFM}
\end{array}
\right.
\end{equation}
In the FM case anisotropy in $f$-system results in an increase of $g_c$ ($%
|G(0)|$ increases with $w$ more rapidly than $|G_{\min }|$), in the 3D AFM
case $g_c$ decreases, and in 2D AFM anisotropy does not influence $g_c$. The
effective coupling constant $g^{*}=g_{ef}(0)$ remains finite at $%
g\rightarrow g_c-0$ and tends to 
\begin{equation}
g_c^{*}=\mu /\tanh (\mu \Delta ).
\end{equation}
The effective (renormalized by spin dynamics) Kondo temperature $T_K^{*}$ is
determined by the condition $1/g_{ef}(-T_K^{*})=0$ and satisfies the
equation 
\begin{equation}
-G(-T_K^{*})=%
\mathop{\rm arctanh}
(\mu /g)
\end{equation}
Due to the minimum, $T_K^{*}$ is also finite at $g\rightarrow
g_c+0,T_K^{*c}=|C_{\min }|.$

\section{Effects of the $f$-system anisotropy on scaling behavior}

Now we treat the physically real case $N=2$ with the anisotropy being
present in the $f$-system only. To this end we have to consider the full
scaling equations for $\mu =0,w\neq 0.$ The most important circumstance to
be taken into account is the renormalization of the magnon frequencies.
Writing down the first integrals of the system (\ref{gl}), (\ref{wl}) 
\begin{eqnarray}
g_{ef}(C)+(2/a)\ln \overline{\omega }_{ex}(C) &=&\text{const}  \label{int} \\
g_{ef}(C)+(2/b)\ln \omega _0(C) &=&\text{const}  \nonumber
\end{eqnarray}
results in 
\begin{eqnarray}
\overline{\omega }_{ex}(C) &=&\overline{\omega }_{ex}\exp (-a[g_{ef}(C)-g]/2)
\label{w+g} \\
\omega _0(C) &=&\omega _0\exp (-b[g_{ef}(C)-g]/2)  \nonumber
\end{eqnarray}
As follows from (\ref{wl}), (\ref{sl}) 
\begin{equation}
\left( \frac{\overline{\omega }_{ex}(C)}{\overline{\omega }_{ex}}\right)
^{1/a}=\left( \frac{\omega _0(C)}{\omega _0}\right) ^{1/b}=\frac{\overline{S}%
_{ef}(C)}S  \label{wws}
\end{equation}
Substituting (\ref{w+g}) into (\ref{gl}) we obtain 
\begin{equation}
\partial (1/g_{ef})/\partial \xi =-\eta (\exp (\xi -\lambda
_{ex}-a[g_{ef}-g]/2),w\exp (\xi -\lambda _{ex}-b[g_{ef}-g]/2))  \label{gpsi}
\end{equation}
where 
\[
\xi =\ln |D/C|,\,\lambda _{ex}=\ln (D/\overline{\omega }_{ex})\gg 1,w=\omega
_0/\overline{\omega }_{ex}.
\]
For an antiferromagnet in the nearest-neighbor approximation we have $a=b$
and the equation (\ref{gpsi}) takes the form 
\begin{eqnarray}
\partial (1/g_{ef})/\partial \xi  &=&-\Psi (\lambda _{ex}-\xi
+a[g_{ef}-g]/2))  \label{gsi} \\
\Psi (\xi ) &=&\eta ^{AFM}(e^{-\xi },w\,e^{-\xi })  \nonumber
\end{eqnarray}

For finite values of $N$ the singularities of the scaling functions, that
occur in the magnetically ordered phases, play the crucial role due to
peculiar properies of the differential equation (\ref{gpsi}). In particular,
one can prove\cite{kondo} that $g_{ef}$ diverges at some $\xi $ at
arbitrarily small $g$ (i.e. $g_c=0$) unless the singularity cutoff is
introduced. To make the value of $g_c$ finite one has to cut the
singularities. This may be performed by introducing small imaginary parts,
i.e. by replacing in (\ref{eta1}),(\ref{eta3}) 
\begin{eqnarray}
\ \ \ \ \ln |1-x| &\rightarrow &\text{Re}\ln [1-x(1+i\delta )]=\frac 12\ln
[(1-x)^2+\delta ^2x^2],  \label{cut} \\
\ \ \ (1-x)^{-1/2}\theta (1-x) &\rightarrow &\{[(1-x)^2+\delta
^2x^2]^{1/2}+1-x]/2\}^{1/2}/[(1-x)^2+\delta ^2x^2]^{1/2}  \nonumber
\end{eqnarray}
The $x$-dependence of the cutoff parameter can be in principle neglected, as
in Ref.\cite{kondo}, since it does not influence appreciably numerical
results (since $\delta $ is important at $x=1$ only); however, this
dependence is needed to pass correctly to the limit $\omega _0\rightarrow 0$%
). As one can see from (\ref{etafm}), the value of $\delta $ should be
determined by the magnon damping at $q=|{\bf k-k}^{\prime }|\simeq 2k_F$.
This damping is due to both exchange and relativistic interactions.
Hereafter we put in numerical calculations $\delta =1/100.$ We accept also $%
\lambda =\ln (D/\overline{\omega })=5.$

The dependences $g_{ef}(\xi )$ for 3D FM and AFM phases according to (\ref
{gpsi}), (\ref{gsi}) at $g$ close to the critical value are shown in Fig.3.
As well as for the isotropic case\cite{kondo}, there occur large intervals
of a non-Ferm-liquid (NFL) behavior where magnon spectrum becomes soft. In
this regime, the relation between the arguments of the scaling function $%
\eta (x,y)$ in (\ref{gpsi}) is fixed by the singularity point $|C|=\overline{%
\omega }(C)$ (see Appendix, $\overline{\omega }(\xi )$ is defined in the
same way as in (\ref{wtot})) or by the condition 
\begin{equation}
\ln [\overline{\omega }/\overline{\omega }(\xi )]=\xi -\lambda  \label{ll}
\end{equation}
After substituting\ (\ref{w+g}) into (\ref{ll}) we see that the dependence $%
g_{ef}(\xi )$ is linear in $\xi $ only in the case $a=b$ where 
\begin{equation}
g_{ef}(\xi )-g\simeq 2(\xi -\lambda )/a  \label{lin}
\end{equation}
Thus for the FM phase the dependence $g_{ef}(\xi )$ in the NFL region is
different from the isotropic case.

In the 2D AFM case the anisotropy does not practically influence the
behavior $g_{ef}(C)$ because of strong singularity of the scaling function.

The dependences $1/g^{*}(g)$ and $\xi ^{*}(g)$ ( $\xi _{}^{*}=\ln (D/T_K^{*})
$) for a 3D anisotropic ferromagnet are shown in Fig.4. These dependences
are more similar to those in the isotropic antiferromagnet rather than
isotropic ferromagnet (see Fig.5 of Ref.\cite{kondo}). In particular, a wide
plateau with $\xi ^{*}(g)\simeq $ $\widetilde{\xi }_c^{*}$ can be seen in
Fig.4, whereas in the isotropic ferromagnet $\xi ^{*}(g\rightarrow g_c+0)$
increases in a not too narrow region. The difference is connected with the
absence of the scaling function maximum in the latter case. Thus the
anisotropy makes the dependence the effective Kondo temperature on the bare
coupling parameter still weaker in comparison with the isotropic case. Due
to the minimum of the function $1/g_{ef}(\xi ),$ $g^{*}$ remains finite at $%
g\rightarrow g_c-0,$ as well as in the limit $N\rightarrow \infty
,\,g^{*}\rightarrow g_c^{*}=1/\Delta $ with $\Delta $ given by (\ref{delta}%
). It should be noted that the function$1/g^{*}(g)$ (Fig.4) approaches $%
\Delta $ at very small $|g-g_c|$ which are practically unreachable.

On the other hand, strictly speaking, $\xi ^{*}(g)$ diverges at $%
g\rightarrow g_c+0.$ The increase of $\xi ^{*}(g)$ takes place also in an
extremely narrow region (of order of 10$^{-4}$-10$^{-3}$) only and is not
shown in Fig.4. As demonstrate numerical calculations, in this region we
have 
\begin{equation}
\xi _{}^{*}(g)-\widetilde{\xi }_c^{*}\simeq -\gamma \ln
[(g-g_c)/g],T_K^{*}\sim (g-g_c)^\gamma  \label{crit}
\end{equation}
The ``critical exponents'' $\gamma $ turn out to be non-universal,
decreasing with increasing $w,$ i.e. the minimum depth; for the 3D ferro-
and antiferromagnets with $w=0.3$ we have $\gamma \simeq 0.2$ and $\gamma
\simeq 0.05$ (the corresponding values in the isotropic case are $\gamma
=1/2 $ and $\gamma \simeq 0.1$ respectively\cite{kondo}).

A comparison of the critical parameter values in the isotropic and
anisotropic cases is presented in the Table 1. One can see that for $N=2$
the anisotropy results in a decrease of the critical value $g_c$ in all the
cases, unlike the large-$N$ limit. This decrease is more appreciable in the
FM phase where linear terms in the anisotropy parameter enter the equations.

Table 1. The critical values $g_c\,$ and $\xi _c^{*}$ for different magnetic
phases in the isotropic and anisotropic cases at $N=\infty $ (from
analytical results, see Sect.4) and $N=2$ (from numerical solution). The
parameter values are $\lambda =5,$ $\alpha =1/2,\,\alpha ^{\prime }=0.$ For $%
N=2\,,$ the cutoff $\delta =1/100$ is used and the ``critical value'' $%
\widetilde{\xi }_c^{*}$ is estimated from the plateau in the dependence $\xi
^{*}(g)$ (see the discussion in the text).

\begin{tabular}{|llllll|}
\hline
\multicolumn{1}{|l|}{} & \multicolumn{1}{l|}{$w$} & \multicolumn{1}{l|}{} & 
\multicolumn{1}{l|}{FM} & \multicolumn{1}{l|}{3D AFM} & 2D AFM \\ \hline
\multicolumn{1}{|l|}{$N\to \infty $} & \multicolumn{1}{l|}{0} & 
\multicolumn{1}{l|}{$g_c$} & \multicolumn{1}{l|}{0.167} & 
\multicolumn{1}{l|}{0.171} & 0.176 \\ \hline
\multicolumn{1}{|l|}{} & \multicolumn{1}{l|}{0.3} & \multicolumn{1}{l|}{} & 
\multicolumn{1}{l|}{0.169} & \multicolumn{1}{l|}{0.170} & 0.176 \\ \hline
\multicolumn{1}{|l|}{} & \multicolumn{1}{l|}{0} & \multicolumn{1}{l|}{$\xi
_c^{*}$} & \multicolumn{1}{l|}{$\infty $} & \multicolumn{1}{l|}{5.35} & 5 \\ 
\hline
\multicolumn{1}{|l|}{} & \multicolumn{1}{l|}{0.3} & \multicolumn{1}{l|}{} & 
\multicolumn{1}{l|}{5.73} & \multicolumn{1}{l|}{5.31} & 5 \\ \hline
\multicolumn{1}{|l|}{$N=2$} & \multicolumn{1}{l|}{0} & \multicolumn{1}{l|}{$%
g_c$} & \multicolumn{1}{l|}{0.139} & \multicolumn{1}{l|}{0.132} & 0.127 \\ 
\hline
\multicolumn{1}{|l|}{} & \multicolumn{1}{l|}{0.3} & \multicolumn{1}{l|}{} & 
\multicolumn{1}{l|}{0.133} & \multicolumn{1}{l|}{0.130} & 0.127 \\ \hline
\multicolumn{1}{|l|}{} & \multicolumn{1}{l|}{0} & \multicolumn{1}{l|}{$%
\widetilde{\xi }_c^{*}$} & \multicolumn{1}{l|}{6.13} & \multicolumn{1}{l|}{
6.07} & 6.07 \\ \hline
\multicolumn{1}{|l|}{} & \multicolumn{1}{l|}{0.3} & \multicolumn{1}{l|}{} & 
\multicolumn{1}{l|}{6.24} & \multicolumn{1}{l|}{6.04} & 6.06 \\ \hline
\end{tabular}

\noindent The value of $\widetilde{\xi }_c^{*}$ decreases with anisotropy in
the AFM case but increases in the FM case. This is explained by that in FM $%
\omega _0$ is renormalized stronger than $\overline{\omega }_{ex}$, so that $%
\overline{\omega }(\xi )$ tends to zero more rapidly. As demonstrate
numerical calculations, $\widetilde{\xi }_c^{*}$ decreases with $w$ in the
AFM case too for sufficiently large $\alpha ^{\prime }$.

The experimentally observable quantities can be obtained by using the
formulas 
\begin{equation}
T_K^{*}=D\exp (-\xi ^{*})
\end{equation}
($g>g_c$) and 
\begin{eqnarray}
S^{*} &=&\overline{S}_{ef}(C=0)=S\exp (-[g^{*}-g]/2)  \nonumber \\
\overline{\omega }_{ex}^{*} &=&\overline{\omega }_{ex}(C=0)=\overline{\omega 
}_{ex}\exp (-a[g^{*}-g]/2)  \label{sww0} \\
\omega _0^{*} &=&\omega _0(C=0)=\omega _0\exp (-b[g^{*}-g]/2)  \nonumber
\end{eqnarray}
($g<g_c$). In particular, we obtain the relation 
\begin{equation}
\omega _0^{*}/\omega _0=\left\{ 
\begin{array}{cc}
(S^{*}/S)^2 & \text{FM} \\ 
S^{*}/S & \text{AFM}
\end{array}
\right.
\end{equation}
The quantities (\ref{sww0}) are finite at $g\rightarrow g_c-0.$ However, as
follows from the standard scaling treatment of the phase transition, in some
region we have the law 
\begin{equation}
\overline{\omega }_{}^{*}=\overline{\omega }(C=0)\sim (g_c-g)^\gamma
\end{equation}
Renormalization of relative anisotropy parameter is given by 
\begin{equation}
w(C)=\frac{\omega _0(C)}{\overline{\omega }_{ex}(C)}=\exp (-\frac{b-a}2%
[g_{ef}(C)-g])  \label{renw}
\end{equation}
This is shown in Fig.5. The corresponding temperature dependence can be
obtained by the replacement $|C|\rightarrow T.$

\section{Effects of anisotropic $s-f$ coupling}

Strictly speaking, in the case where $\mu \neq 0$ the full system of scaling
equations cannot be simplified. However, a simple analytical treatment is
possible in the case $\mu \ll g$ which is physically real for magnetic
systems. Under this condition we can expand 
\begin{equation}
g_{ef}^{\perp }=\sqrt{g_{ef}^2-\mu ^2}\simeq g_{ef}-\frac 12\mu ^2/g_{ef}
\end{equation}
Provided that the expansion holds at $\xi =0,$ this will hold with
increasing $g_{ef}$ too. Then we obtain from (\ref{gl}),(\ref{wl}),(\ref{w0}%
) the integrals of motion 
\begin{eqnarray}
g_{ef}(C)-\tau \mu ^2/g_{ef}(C)+(2/a)\ln \overline{\omega }_{ex}(C) &\simeq &%
\text{const, }\tau =\left\{ 
\begin{array}{cc}
1/2 & \text{FM} \\ 
-\alpha  & \text{AFM}
\end{array}
\right.   \label{gwmu} \\
g_{ef}(C)-\theta \mu ^2/g_{ef}(C)+(2/b)\ln \omega _0(C) &\simeq &\text{%
const, }\theta =\left\{ 
\begin{array}{cc}
1/2 & \text{FM} \\ 
-(1-\alpha )s & \text{AFM}
\end{array}
\right.   \label{gw0mu}
\end{eqnarray}
so that 
\begin{eqnarray}
\overline{\omega }_{ex}(C) &=&\overline{\omega }_{ex}\exp
(-a[g_{ef}(C)-g-\tau \mu ^2(1/g_{ef}(C)-1/g)]/2)  \label{wmu1} \\
\omega _0(C) &=&\overline{\omega }_{ex}\exp (-b[g_{ef}(C)-g-\theta \mu
^2(1/g_{ef}(C)-1/g)]/2)  \label{wmu2}
\end{eqnarray}
On substituting (\ref{wmu1}),(\ref{wmu2}) into (\ref{gmu}) we obtain the
closed equation 
\begin{eqnarray}
\partial g_{ef}/\partial \xi  &=&-(g_{ef}^2-\mu )\eta (\exp (\xi -\lambda
_{ex}-a[g_{ef}-g-\tau \mu ^2(1/g_{ef}-1/g)]/2),  \nonumber \\
&&\ \ \ \ \ \ \ w\exp (\xi -\lambda _{ex}-b[g_{ef}-g-\theta \mu
^2(1/g_{ef}-1/g)]/2))  \label{gpsimu}
\end{eqnarray}
Presence of the terms, that are proportional to $\mu ,$ in the scaling
function arguments in (\ref{gpsimu}), results in a weak smearing of the
linear dependence $g_{ef}(\xi )$ in the NFL regime even for $a=b$.

In the FM case, to accuracy accepted, the expressions (\ref{wmu1}),(\ref
{wmu2}) can be represented as 
\begin{eqnarray}
\overline{\omega }_{ex}(C) &=&\overline{\omega }_{ex}\exp (-a[g_{ef}^{\perp
}(C)-g_{\perp }]/2) \\
\omega _0(C) &=&\omega _0\exp (-b[g_{ef}^{\perp }(C)-g_{\perp }]/2)
\end{eqnarray}
Further, as follows from (\ref{sl}), in all the cases 
\begin{equation}
\overline{S}_{ef}(C)=S\exp (-[g_{ef}(C)-g]/2)
\end{equation}
Thus the first of relations (\ref{wws}) is violated in AFM case, and the
second relation in both FM and AFM cases.

However, the violation owing the anisotropic $s-f$ coupling is weak:
renormalization of $\omega _0$ for AFM, which is most appreciable, yields in
the exponent of (\ref{wmu2}) the quantity of the order of $\mu ^2s/g\propto
g $ only. Thus the most important effect of $\mu $ is the deformation of
scaling trajectories at not too large values of $\xi ,$ which was considered
in Sect.4 (Fig.1), and main corrections to $g_c$ are described by (\ref{gcmu}%
).

The case of not small $\mu $ (which can be relevant for pseudospin systems%
\cite{ikt1}) can be investigated by numerical solving the equations (\ref{gl}%
)- (\ref{w0}). The results are shown in Fig.6. One can see that the
anisotropy of $s-f$ coupling leads to an increase of the $g_c$ values in
comparison with the isotropic case (cf. Table 1). This is due to both
decrease of $g_{\bot }$ and more weak renormalization of $\overline{\omega }%
_{ex}$ according to (\ref{wl}).

\section{Conclusions}

In the present paper we have generalized the scaling treatment of Ref.\cite
{kondo} by including the anisotropy in both the $s-f$ coupling and $f$%
-system itself. We have demonstrated that the magnetic anisotropy modifies
considerably the scaling behavior in the Kondo lattice problem.

In all the cases, the anisotropy in the $f$-subsystem (the gap in the magnon
spectrum) results in occurrence of a non-monotonous dependence (of a
maximum) of the effective coupling constant $g_{ef}(\xi )$. This prevents
the increase of $\xi ^{*}(g)$ at $g\rightarrow g_c+0$, which becomes
practically not observable even for a ferromagnet (Fig.4), unlike the
isotropic case. The dependence of the effective Kondo temperature on the
bare $s-f$ coupling parameter becomes weaker in the presence of the
anisotropy. Further, the minimum in the dependence $1/g_{ef}(\xi )$ results
in that $g^{*}$ (and, consequently, $\overline{\omega }_{ex}^{*},\,\omega
_0^{*}$, and $S^{*}$) are always finite at $g\rightarrow g_c-0$. As for
quantitative changes, the anisotropy favors a non-magnetic Kondo state (the
critical values of bare coupling constant decrease) The critical region of
magnetic instability becomes more narrow (especially in the FM case), so
that the non-Fermi-liquid behavior is suppressed. At the same time,
anisotropic $s-f$ coupling influences noticeably the form of the scaling
trajectory for small $\xi ,$ but becomes not important with increasing $%
g_{ef}(\xi ).$

Owing to a more rapid Kondo renormalization of the gap, the system tends to
``isotropic'' behavior of $g_{ef}(\xi )$ at approaching the strong coupling
regime. Such a renormalization (\ref{renw}) may be important for analysis of
the spin excitation spectrum in anomalous $f$-systems. One can expect that
the observable renormalized spectrum gap in these systems is relatively
small and strongly temperature-dependent in comparison with the ``usual'' $f$%
-electron magnets. As demonstrate our calculations, next-nearest-neighbor
exchange interactions in AFM phase are important for this effect. The
dependence $\omega _0(T)$ can be investigated not only by the neutron
scattering, but also by simple methods like the ferromagnetic resonance.

The change of the critical exponents of the phase transition at $%
g\rightarrow g_c$ with changing the bare anisotropy parameter $w$ turns out
to be very strong. It is interesting that their values depend continuously
on the anisotropy and are non-universal (possibly, higher order
contributions to the scaling equations will change this result; this
question needs further investigations). It should be noted that, unlike the
one-mpurity Kondo problem (where the ``phase transition'', connected with
disappearance of the {\it local} moment, exists for $N=\infty $ only, and a
crossover takes place for finite $N$, see Ref.\cite{Col}), the phase
transition in the Kondo lattices is physically real, being a
magnetic-nonmagnetic transition. The situation is similar to the onset of
magnetism in itinerant electron systems. Of course, one has to bear in mind
that the treatment of this transition within the lowest-order scaling may be
only qualitatively correct, and a more detailed (e.g., numerical
renormalization group) consideration is needed to describe the crossover
region.

Similar to Ref.\cite{kondo}, at approaching the critical value of magnetic
instability $g_c,$ the transition to an ``incoherent'' regime, where
non-spin-wave excitations of $f$-system play the dominant role, should be
considered. In this regime, the minimum of $1/g_{ef}(\xi )$ can be
suppressed. However, the use of the model ``paramagnetic'' scaling function
for describing the incoherent contribution (as in Ref.\cite{kondo}) seems to
be unreasonable in the presence of the anisotropy, since spin dynamics in
the paramagnetic state of $f$-systems in a strong crystal field is rather
complicated\cite{FLow,Maekawa}.

The work was supported in part by Grant 96-02-16000 from the Russian Basic
Research Foundation.

\appendix

\section{Appendix. Scaling functions in anisotropic magnets}

Using the long-wave approximations $\omega _{ex}^{FM}({\bf q})\sim q^2,$ $%
\omega _{ex}^{AFM}({\bf q})\sim q$ in the whole Brillouin zone (which is
justified, e.g., at small $k_F$)$,$ we get from (\ref{etafm}) 
\begin{equation}
\eta ^{FM}(x,y)=\frac 1{2x}\ln \left| \frac{1+x+y}{1-x-y}\frac{1-y}{1+y}%
\right|  \label{eta1}
\end{equation}
where $\overline{\omega }_{ex}=\omega _{ex}(2k_F).$ For an antiferromagnet
we derive 
\begin{equation}
\eta ^{AFM}(x,y)=\frac 1{1-y^2}\eta ^{AFM}(\frac x{\sqrt{1-y^2}})
\end{equation}
where $\eta ^{AFM}(x)$ is the corresponding scaling function in the
isotropic case,

\begin{eqnarray}
\eta ^{AFM}(x)= 
{-x^{-2}\ln |1-x^2|,\,d=3 \atopwithdelims\{. (1-x^2)^{-1/2}\theta (1-x^2),\,d=2}
\end{eqnarray}
$\theta (x)$ being the step function. Then we have 
\begin{equation}
\eta ^{AFM}(x,y)= 
{\frac 1{x^2}\ln \left| \frac{1-y^2}{1-x^2-y^2}\right| ,\,d=3 \atopwithdelims\{. \frac{\theta (1-x^2-y^2)-\theta (y^2-1)}{|1-y^2|^{1/2}|1-x^2-y^2|^{1/2}},\,d=2}
\label{eta3}
\end{equation}
One can see that the logarithmic singularities of the functions $\eta
^{FM}(x)=$ $\eta ^{FM}(x,0)$ and $\eta ^{AFM}(x)$ at $x=1$ are shifted to
the points $x+y=1$ (FM) and $x^2+y^2=1$ (AFM), i.e. $|C|=\overline{\omega }$
with $\overline{\omega }$ defined by (\ref{wtot}). Besides that, the
anisotropy results in occurrence of the second singularity at $y=1$ (i.e., $%
|C|=\omega _0$).

Presence of such singularities is a general property which does not depend
on the spectrum model\cite{kondo}. The functions $\eta ^{FM}(x,y)$ and $\eta
^{AFM}(x,y)$ ($d=3)$ change their sign at $y(x+y)=1,$ i.e. 
\begin{equation}
|C|=|C_{\min }|=\sqrt{\omega _0(\omega _0+\overline{\omega }_{ex})}
\label{min1}
\end{equation}
and $x^2+2y^2=2,$ i.e. 
\begin{equation}
|C_{\min }|=\sqrt{\omega _0^2+\overline{\omega }_{ex}^2/2}  \label{min2}
\end{equation}
respectively. For $d=2$ the function $\eta ^{AFM}(x)$ has strong square-root
singularities at $x^2+y^2=1$ and $y=1,$ and vanishes in the interval $%
1-x^2<y^2<1,$ i.e. 
\begin{equation}
|C_{\min }|=\sqrt{\omega _0^2+\overline{\omega }_{ex}^2}>|C|>\omega _0,
\label{min3}
\end{equation}
changing its sign after passing this interval (but a smooth contribution
occurs for more realistic models of magnon spectrum).

Note that in the limit of strong magnetic anisotropy $\omega _0/\overline{%
\omega }_{ex}\rightarrow \infty $ the singularity at $y\rightarrow 1$
becomes very strong: 
\begin{equation}
\eta ^0(y)=(1-y^2)^{-1}  \label{sing}
\end{equation}
It should be noted that the influence of magnetic anisotropy on spin
dynamics can be also considered for the paramagnetic phase (i.e., for the
problem of the local moment formation). In this case the singularity (\ref
{sing}) also becomes weaker. If we accept, as in Ref.\cite{kondo}, the
``spin diffusion'' approximation with the spin spectral density 
\begin{equation}
{\cal J}_{{\bf q}}(\omega )=\frac 1\pi \frac{{\cal D}q^2}{(\omega -\omega
_0)^2+({\cal D}q^2)^2}  \label{JD}
\end{equation}
(${\cal D}$ is the spin diffusion constant, $\overline{\omega }_{ex}=4{\cal D%
}k_F^2$) we obtain 
\begin{equation}
\eta ^{PM}(x,y)=\frac 1{2x}\left( \arctan \frac x{1-y}+\arctan \frac x{1+y}%
\right)
\end{equation}
This function has a finite jump at $|C|=\omega _0.$ Of course, this
approximation can be hardly justified for real $f$-systems. Therefore we do
not present concrete calculation results for the PM case. However, one can
expect that qualitative effects of anisotropy are similar to those in the
magnetically ordered phases.

{\sc Figure captions}

Fig.1. The dependence $1/g_{ef}$ on $\xi =\ln |D/C|$ in the case of
anisotropic $s-f$ coupling for a 3D antiferromagnet (upper solid line), 2D
antiferromagnet (short-dashed line) and 3D ferromagnet (long-dashed line).
The parameters are $\lambda =\ln (D/\overline{\omega })=5,\,g=0.15,$ $\mu
=0.1$. The lower solid line shows the curve for the 3D antiferromagnet with $%
\mu =0$.

Fig.2. The dependence $1/g_{ef}$ on $\xi =\ln |D/C|$ in the presence of
anisotropy in $f$-system for a 3D antiferromagnet (solid line), 2D
antiferromagnet (short-dashed line) and 3D ferromagnet (long-dashed line).
The parameters are $\lambda =5,\,g=0.15,$ $w=0.3.$

Fig.3. The scaling trajectories $g_{ef}(\xi )$ in an anisotropic ferromagnet
for $g=0.1333>g_c$ (upper solid line) and $g=0.1332<g_c$ (lower solid line),
and 3D antiferromagnet for $g=0.1302>g_c$ (upper dashed line) and $%
g=0.1301<g_c$ (lower dashed line), $w=0.3$.

Fig.4. The dependences $1/g^{*}(g)$ for $g<g_c$ and $\xi ^{*}(g)-\lambda $
for $g>g_c$ in an anisotropic ferromagnet with $\,\lambda =5,\,\alpha
=1/2,\,\,w=0.3,$ $\delta =1/100$. The dashed line is the curve $1/g-\lambda .
$

Fig.5. The scaling behavior of the effective anisotropy parameter $%
w_{ef}(\xi )=\omega _0(\xi )/\overline{\omega }_{ex}(\xi )$ for an
anisotropic ferromagnet with the same parameters as in Fig.2 (solid lines)
and antiferromagnet with $\alpha ^{\prime }=0.4,$ $g=0.1375>g_c$ (upper
dashed line) and $g=0.1374<g_c$ (lower dashed line).

Fig.6. The dependences $1/g_{ef}(\xi )$ in the case of anisotropic $s-f$
coupling ($\mu =0.1,w=0$) for a 3D ferromagnet with $g=0.1619>g_c$ (lower
solid line) and $g=0.1618<g_c$ (upper solid line) and a 3D antiferromagnet
with $g=0.1563>g_c$ (lower dashed line) and $g=0.1562<g_c$ (upper dashed
line).

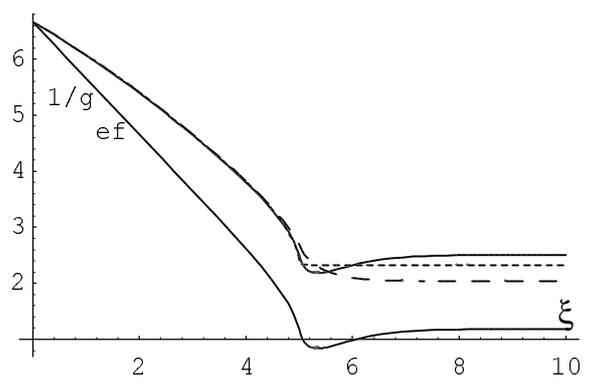
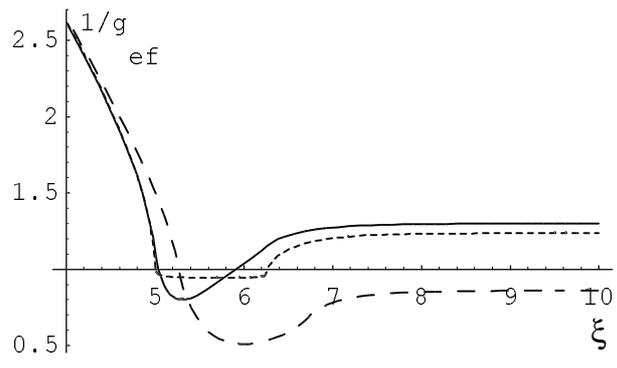
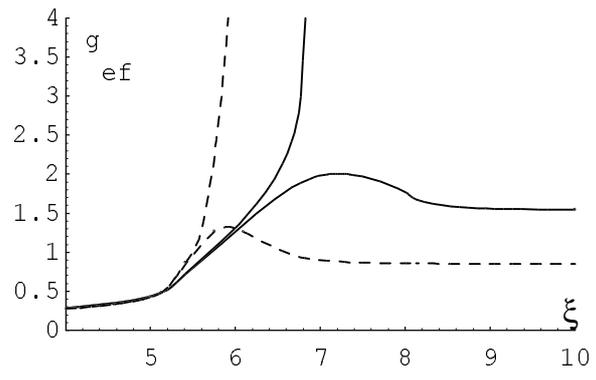
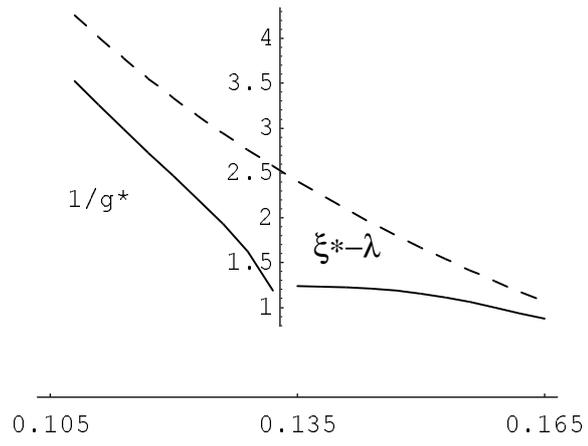
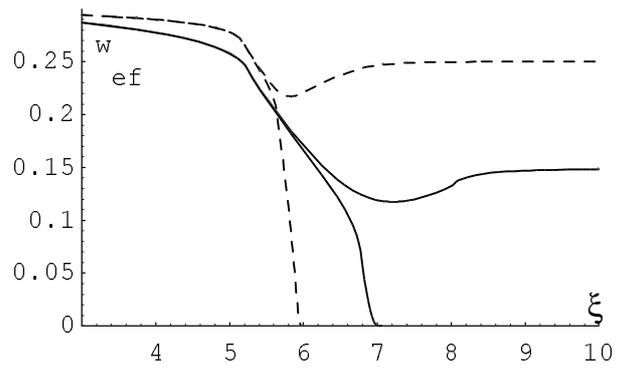
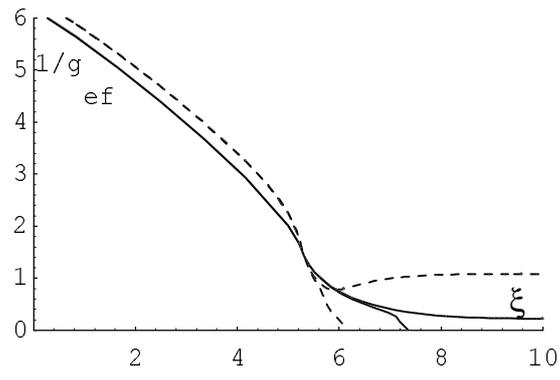
\end{document}